\documentclass[pre,twocolumn,aps,superscriptaddress,longbibliography]{revtex4}
\usepackage[version=3]{mhchem} 
\usepackage[T1]{fontenc}       
\usepackage{epsfig,amsmath,amssymb,graphicx,color,calc,epstopdf}
\usepackage{inputenc}

\def\to{\rightarrow}

\newcommand{\br}{\mathbf{r}}
\newcommand{\ud}{\mathrm{d}}

\newcommand{\beq}{\begin{equation}}
\newcommand{\eeq}{\end{equation}}
\newcommand{\ba}{\begin{align}}
\newcommand{\ea}{\end{align}}

\makeatletter 
\@addtoreset{equation}{section}
\makeatother  

\newcommand{\zyn}[1]{{{#1}}}
\newcommand{\zy}[1]{{{#1}}}

\begin{document}
\title{Equilibrium Phase Behavior of the Square-Well Linear Microphase-Forming Model}
\author{Yuan Zhuang}
\affiliation{Department of Chemistry, Duke University, Durham,
North Carolina 27708, USA}
\author{Patrick Charbonneau}
\affiliation{Department of Chemistry, Duke University, Durham,
North Carolina 27708, USA}
\affiliation{Department of Physics, Duke University, Durham,
North Carolina 27708, USA}
\email{patrick.charbonneau@duke.edu}

\begin{abstract}
We have recently developed a simulation approach to calculate the equilibrium phase diagram of particle-based microphase former. Here, this approach is used to calculate the phase behavior of the square--well--linear model \zy{for different strengths and ranges of the linear long--range repulsive component}. \zy{The results are compared with various theoretical predictions for microphase formation. The analysis further allows us to better understand the mechanism for microphase formation in colloidal suspensions.}
\end{abstract}

\maketitle
\section{Introduction}
In simple liquids, gas--liquid coexistence results from particles attracting one another.
When a sufficiently strong and long-range repulsion frustrates this
attraction, however, microphases supersede the condensed phase. Interestingly, the formation of these mesoscale structures is nearly as \zy{common} as the bulk behavior it replaces. Microphases are indeed observed in systems as diverse as magnetic alloys~\cite{Seul1995}, Langmuir films~\cite{Keller1986}, and protein solutions~\cite{Stradner2004}, irrespective of the physical origin of the competing short-range attractive and long-range repulsive (SALR) interaction~\cite{Seul1995,Ciach2013}.

Microphases can also \zy{become ordered,} \zy{resulting in structures that are} both elegant and useful. For instance, block
copolymers~\cite{Bates1990,Bates1999,Kim2010}--wherein chain connectivity \zy{frustrates the immiscibility of the different chain components}--can form a
rich array of periodic morphologies, such as clusters, lamellae,
and gyroid~\cite{Leibler1980,Matsen1996}, as well as exotic
structures, such as zigzag~\cite{Lee2001} and $O^{70}$~\cite{Tyler2005} phases.
The experimental controllability and reproducibility of some of these structures further enable their industrial uses for drug
delivery~\cite{Kataoka2001,Rosler2001} nanoscale
patterning~\cite{Li2006,Krishnamoorthy2006}, and lithography~\cite{Hawker2005,Tang2008}, among others.

From a theoretical viewpoint, our understanding of microphase assembly is \zy{somewhat} uneven. Insights from field theory,
density-functional theory~(DFT),
self-consistent field theory~(SCFT), random-phase
approximation~(RPA) and
others~\cite{Leibler1980,Bates1990,Matsen1994,Matsen1996,Sear1999,Liu2005, Imperio2004,Archer2007, Brazovskii1975,Seul1995,Archer2008, Chacko2015} have been collated to describe various aspects of the thermodynamics and \zy{mesoscale} structure of microphases.
Some of the more microscopic features of microphase formers, however, remain harder to pinpoint. In particular, the relationship between the ordered microphase regime and the disordered phase that surrounds it has been challenging to \zy{fully} capture. A cluster fluid and a percolating (gel-like) fluid have been argued to either surround or substitute for the microphase regime; \zy{but} whether the physical origin of these disordered features is a question of equilibrium~\cite{Toledano2009,Liu2005} or out-of-equilibrium dynamics~\cite{deCandia2006,Tarzia2007,Charbonneau2007,Schmalian2000,Geissler2004}, or more akin to defect annealing ~\cite{Zhang2006,deCandia2011,DelGado2010} can be difficult to tease out. In addition to being an interesting physical question in its own right, it also muddles the interpretation of certain experiments. Suspensions of colloidal particles with SALR interactions, for instance, have \zy{been observed to from clusters and gels, but} assemblies of periodic microphases \zy{have not yet been obtained}~\cite{Stradner2004,Jordan2014,Campbell2005,Klix2010,Zhang2012b,Klix2013}.  In order to understand microphase assembly, the relationship between equilibrium statics and dynamics \zy{must be deconvoluted}.

In this context, dynamical and thermodynamical insights from simulations would be beneficial. The numerical route, however, faces steep obstacles \zy{when periodic} microphases \zy{are involved}. The two main approaches \zy{traditionally used} for delineating phase boundaries \zy{encounter serious difficulties}. First, if one is willing to tolerate the hysteresis due to such a transition would severely caused by the interfacial free energy, slow annealing from high temperatures can sometimes suffice to sketch a phase diagram. The sluggish dynamics of the disordered phase in the vicinity of the microphase regime, however, limits the usefulness of this approach for identifying \zy{that} regime. Second, even hysteresis-free and high-accuracy approaches developed for studying equilibrium phase behavior, such as Widom insertion, Gibbs ensemble Monte Carlo, and Frenkel-Ladd thermodynamic integration~\cite{Frenkel2002}, here \zy{suffer.} The periodicity and occupancy of the mesoscale features can indeed be kept metastable relatively far from equilibrium, for arbitrarily long simulation times~\cite{Zhang2011}. Finite-size effects are thus much more pronounced than for \zy{simpler models}.

We have recently developed an approach to minimize these finite-size effects and identify the equilibrium phase behavior of simple microphase formers using numerical simulations~\cite{Zhuang2016}. This scheme builds on that developed for \zy{surmounting} similar \zy{challenges} in crystals with vacancies~\cite{Swope1992}, cluster crystals~\cite{Mladek2007}, and lattice \zy{SALR} models~\cite{Zhang2011}. Here, we use this method to systematically study a model microphase former and compare the results with various analytical approximations. The plan for the rest of this paper is as follows. We first present a schematic microphase-forming model with a SALR interaction, and then detail the set of numerical tools \zy{used} to examine its phase behavior. \zy{Various} theoretical approximations of the phase behavior of this model are then presented. Finally, we discuss both differences between simulation and theory and the experimental consequences of our results, before briefly concluding.

Before diving into the core of the material, however, let us \zy{first} make a terminological note. While gas-liquid coexistence terminates at a second-order critical point, Brazovskii predicted that the microphase regime terminates at a weakly first-order order-disorder transition (ODT) for systems with isotropic repulsion~\cite{Brazovskii1975}. This behavior contrasts with that observed in systems with anisotropic repulsion, for which the ODT is \zy{then} critical~\cite{Diehl2000}. The transition from gas-liquid critical point to ODT in anisotropic systems is \zy{thus} itself a high-order critical point, a so-called Lifshitz point~\cite{Brazovskii1975,Diehl2000}. Some authors have chosen to mildly abuse the term by using it to also describe the comparable (but noncritical) transition in systems with isotropic repulsion~\cite{Pini2000}. For convenience, we \zy{also} here make this analogy.

\section{Model}
\label{sec:model}
\zy{SALR interactions in colloidal suspensions are typically due to short-range depletion attraction competing with long-range screened electrostatic repulsion~\cite{Stradner2004,Campbell2005}. The precise physical origin of these interactions should not, however, play a large role on the system behavior. Critical Casimir attraction~\cite{Nguyen2013} and magnetic repulsion~\cite{Meuller2014}, for instance, might also give rise to similar structures, although no such experimental system has yet been considered.}

\zy{Here, we model the generic features of colloidal SALR interactions with a schematic square--well--linear (SWL) model. This model has a radially symmetric pair interaction potential, resulting in the total potential energy for a configuration $\mathbf{r}^N$ with $N$ particles to be given by}
\begin{equation}
U(\mathbf{r}^N)=\sum_{i<j} u(r_{ij}),
\end{equation}
where the SALR pair interaction potential can itself be subdivided in two, $u(r)=u_{\mathrm{HS}}(r)+u_{\mathrm{SALR}}(r)$. The first term contains the hard sphere (HS) volume exclusion
\begin{equation}
u_{\mathrm{HS}}(r)=
\begin{cases} \infty &\mbox{if } r \leq \sigma \\
0 & \mbox{if } r > \sigma \end{cases}
\end{equation}
for spheres of diameter $\sigma$, where $\sigma$ sets the unit of length. The second term \zy{is} the SALR contribution proper,
\begin{equation}\label{eq:swl}
u_{\mathrm{SALR}}(r)=\begin{cases} -\varepsilon &\mbox{if } r<\lambda\sigma \\
\xi\varepsilon(\kappa-r/\sigma) & \mbox{if } \lambda\sigma<r<\kappa\sigma \\
 0 & \mbox{if } r>\kappa\sigma \end{cases},
\end{equation}
comprised of a square--well attraction of \zy{width $\lambda\sigma$ and} strength $\varepsilon>0$--where $\varepsilon$ sets the unit of energy--followed beyond $\lambda\sigma$ by a repulsive ramp of initial amplitude $\xi\varepsilon(\kappa-\lambda)$ that decays linearly with $r$.
In order to remain in a regime where, at small $\xi$, the gas-liquid critical point is stable with respect to the solubility line of the crystal, we chose $\lambda=3/2$~\cite{Noro2000}. Working in this regime (rather than at smaller $\lambda$) and its \zy{SALR} perturbation also has the advantage of \zy{increasing} the reversibility of interparticle \zy{bonding}. Here, we consider the behavior of systems with $\kappa=2\to6$ for $\xi$ both below (simple liquid) and above (microphase former) the  Lifshitz point\zy{,} $\xi_{\mathrm{L}}$.

\section{Simulation Methods}
\label{sec:simulation}
In this section we describe the Monte Carlo (MC) simulation methods used to extract the phase behavior of the SWL model. Note that in general, we run $5-7$ replicates for $10^6$ MC steps at each state point studied. In order to ensure sufficient sampling, different sets of MC moves are used for different regimes. Specifics are provided along with the method descriptions below, but all regimes share a core sampling scheme. Single-particle local moves are \zy{systematically} performed using Metropolis MC with a maximal local displacement tuned, such that the acceptance ratio is kept between $40\%\text{--}60\%$. Two types of nonlocal MC moves are also systematically \zy{performed}: (i) $10\%$ of the moves are system-wide random displacements, and (ii) $10\%$ are aggregation volume biased (AVB) moves, which specifically displace a particle to the surface of another~\cite{Chen2000}.  The AVB ``in'' region is set to the attraction radius, i.e., $1<r<\lambda$, and the  ``out'' region is the rest of the system.

\subsection{Expanded Thermodynamics}

\begin{table*}
	\caption{Measures of occupancy and reference field for the relevant microphase morphologies}
	\label{tab:occ}
	\begin{tabular}{p{0.12\textwidth} p{0.18\textwidth}   p{0.7\textwidth}}
		\hline   \hline
		Morphology &         Measure of occupancy  & Reference field, $\mathcal{F}(\br)$\\
		\hline   \hline
		Lamellar &          Average area density per layer & $\mathcal{F}_0\cos (kz)$ \\   \hline{}
		Cylindrical &     Average line density per cylinder  & $\begin{aligned}\mathcal{F}_0 &\cos (kx)\cos\left[k
		\left(
		\frac{x}{2}+\frac{\sqrt{3}y}{2}
		\right)\right] \cos\left[k
		\left(
		\frac{x}{2}-\frac{\sqrt{3}y}{2}
		\right)\right]\end{aligned}$\\   \hline
		FCC-cluster crystal    & Average number of particles per cluster    & $\mathcal{F}_0 \cos(kx)\cos(ky)\cos(kz)$\\   \hline{}
		
		Double Gyroid &           Average number of particles
		per unit cell.  & $\begin{aligned}&-\mathcal{F}_0 \sin \left[g(x,y,z)^2 \right],
		\mathrm{where\quad}\\ &g(x,y,z)=\cos(kx)\sin(ky)+\cos(ky)\sin(kz) +\cos(kz)\sin(kx)\end{aligned} $ \\   \hline
		\hline{}
	\end{tabular}
\end{table*}

The periodic microphase regime is the most challenging one for which to obtain equilibrium configurations and free energy information. \zy{These equilibrium structures} form a periodic lattice\zy{ that does} not have a fixed occupancy. For a temperature-density state point, one can thus obtain small features that are close to one another, or bigger features that are further apart. \zy{Moreover,} although the thermodynamic ground state has a well-defined average feature size and spacing, the finite-size systems used in simulations can result in the stabilization of metastable states over arbitrarily long times~\cite{Mladek2008}.

An expanded thermodynamics has been developed in order surmount comparable difficulties in crystals with vacancies~\cite{Swope1992} and in multiple-occupancy (cluster) crystals~\cite{Mladek2007,Mladek2008}. \zy{The solution entails adding} to entropy--temperature $ST$, pressure--volume $PV$ and chemical potential--number of particles $\mu N$, a pair of conjugate variables: the lattice occupancy $n_{\mathrm{c}}$ and a chemical potential-like $\mu_{\mathrm{c}}$. The differential form of the Helmholtz free energy is then
\begin{equation}
\ud F_{\mathrm{c}} = -S\ud T-P \ud V+\mu \ud N + \mu_{\mathrm{c}}N\ud n_{\mathrm{c}},
\end{equation}
and at equilibrium, $F(N,V,T)=F_{\mathrm{c}}(N,V,T,n_{\mathrm{c}}^{\mathrm{eq}})$. 
\zy{which corresponds to}
\begin{align}
\left(\frac{\partial F_{\mathrm{c}}}{\partial n_{\mathrm{c}}}\right)_{\rho,T;n_{\mathrm{c}}=n_{\mathrm{c}}^{\mathrm{eq}}}&=0.
\end{align}
\zy{with $\rho=N/V$}. The computational strategy consists of obtaining the free energy of various morphologies for $n_{\mathrm{c}}^{\mathrm{eq}}$, and then use those results along with the free energy results for the disordered phase to delimit the phase diagram of the model by common tangent construction. 

Because it is not a genuine physical quantity, the most \zy{computationally} convenient definition of $n_\mathrm{c}$ for a given system can be chosen at will, \zy{and thus we select different definitions} for each microphase morphology (Table~\ref{tab:occ}). For instance, for the lamellar phase, we use the area number density per lamella, \zy{$\varrho_{\ell}$}, while for cluster crystal phases, we use the average number of particles per cluster, $n_{\mathcal{C}}$.

\subsection{Two-Step Thermodynamic Integration}
\begin{figure*}
	\includegraphics{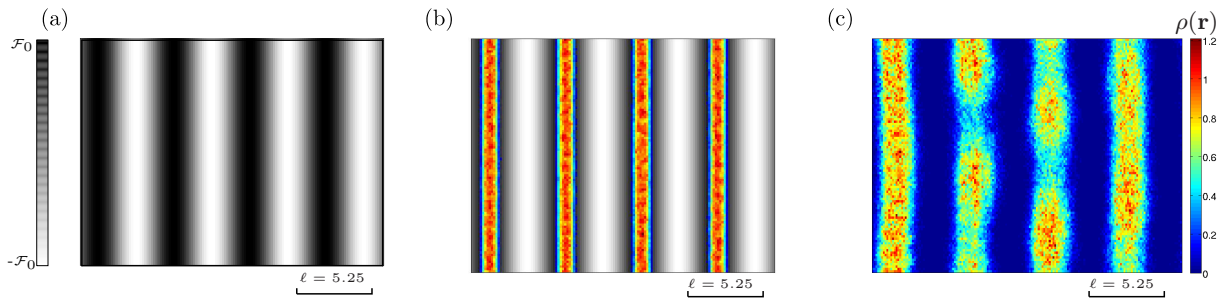}
	\caption{
		Schematic representation of the two-step TI for the lamellar phase at $T=0.3$ and $\rho=0.4$.
		Projections on the $xz$ plane of the number density $\rho(\mathbf{r})$ (coarse-grained over a volume of roughly 1.4 in units of $\sigma^3$)
		and external field profiles $\mathcal{F}(\br)$ for (a) $\rho=0$ with
		field $\mathcal{F}(\br)=\mathcal{F}_0\cos(2\pi z/\ell)$, where
		$\mathcal{F}_0=2$ and $\ell=5.25$, (b) $\rho=0.4$ with
		field and (c) without field.}\label{fig:two_step_ti}
\end{figure*}

In order to obtain the free energy of a given periodic microphase, we devise a two-step thermodynamic integration (TI) protocol, as schematized in Fig.~\ref{fig:two_step_ti}. This approach extends that developed for studying a lattice microphase former~\cite{Zhang2010b} to an off-lattice model with a richer set of microphase morphologies. A key requirement for the success of this scheme is to devise an integration path that connects a system whose free energy is known to the system of interest, while avoiding any first-order phase transition. The hysteresis such a transition would bring would \zy{severely} reduce the accuracy of the method.

We choose as reference state the low-density limit of hard spheres under an external modulated field, $\mathcal{F}(\br)$, whose symmetry is akin to the microphase morphology of interest~(Table~\ref{tab:occ}). \zy{In most cases}, we use combinations of trigonometric functions with a wavevector $k=2\pi/\ell$ that controls the distance $\ell$ between periods of the field. For example, for the lamellar phase, which is periodic in one dimension, we use a simple sinusoidal function $\mathcal{F}(\br)=\mathcal{F}_0\cos (kz)$,
with $k$ depending on the lattice occupancy, $\varrho_\ell=\rho \ell$. The \zy{choice of} field amplitude, $\mathcal{F}_0$, balances two considerations. \zy{On the one hand, it must be} sufficiently strong for the morphology not to melt as the field is turned off (see below), and \zy{on the other hand it must be} sufficiently weak for sampling to remain effective even in the highest-energy parts of the field. In practice, $\mathcal{F}_0=2$--$5$ is found to be a good compromise.

For the double gyroid morphology, which is a bicontinuous phase that emerges from a minimal surface problem~\cite{Podneks1996,Ciach2008}, the field that couples to the particle distribution is a bit more complex. It is chosen to be a sinusoidal function of an approximate description of the minimal surface. \zy{Note that this choice of function softens the edge of the gyroid volume, which facilitates sampling in that region.}

\zy{Overall}, the reference free energy for HS in a field is~\cite{Hansen1990}
\begin{equation}
f_{0,\mathcal{F}}(\beta,\rho)=f^{\mathrm{id}}(\rho)+\int_Ve^{-\beta \mathcal{F}(\br)}\ud\br,
\end{equation}
where $\beta f^{\mathrm{id}}(\rho)=\ln(\rho\Lambda^3)$
is the free energy of an ideal gas at density $\rho$, \zy{with} the thermal de Broglie wave length, $\Lambda$, set to unity, without loss of generality. For some field symmetry, this expression can be further simplified, but parts of the integration must \zy{nonetheless} be completed numerically.

The first TI step brings the system density up to the target $\rho$ under a fixed field, which gives
\begin{equation}
f_{\mathcal{F}}(\beta,\rho)=f_{0,\mathcal{F}}(\beta,\rho)+\int_0^{\rho}\frac{P(\rho')-\rho'/\beta}{\rho'^2}\ud\rho'.
\end{equation}
Note that for the lamellar and cylindrical phases, a virtual harmonic spring
\begin{equation}
U_z = \frac{k_z}{2}(L_z-L_{z_{0}})^2
\end{equation}
with stiffness $k_z$ \zy{is also} applied to the box directions parallel to the mesoscale features. This spring ensures that the final configuration preserves the targeted lattice occupancy, but does not otherwise affect TI.

In order to facilitate the numerical evaluation of the first TI step at low densities, we also compute the first two virial coefficients for HS in a field\zy{~\cite{Hansen1990}}
\begin{align}
&B_{2,\mathcal{F}}(\beta)=\int\!\!\!\int \frac{f_{12}}{Z_{\mathcal{F}}(\beta)} e^{-\beta \left[\mathcal{F}(\br_1) +\mathcal{F}(\br_2)\right]}\ud\br_1\ud\br_2\\
&B_{3,\mathcal{F}}(\beta)=\int\!\!\!\int\!\!\!\int\!\!\! \frac{f_{12}f_{23}f_{13}}{Z_{\mathcal{F}}^2(\beta)} e^{-\beta \left[\mathcal{F}(\br_1) +\mathcal{F}(\br_2)+\mathcal{F}(\br_3)\right]}\ud\br_1\ud\br_2\ud\br_3,
\end{align}
where $f_{ij} = \Theta(|\br_i -\br_j|-1)$ is the HS Mayer function with the Heaviside theta function $\Theta(r)$, and $Z_{\mathcal{F}}(\beta)=\int_{\br}\ud\br e^{\beta\mathcal{F}(\br)}$ is the normalization constant.
Again, these expressions can be further simplified for specific field symmetries, but the integration must \zy{nonetheless} be completed numerically.

Because the first TI step is performed \zy{under a} constant pressure \zy{MC scheme}, we include logarithmic volume moves for a fraction $1/N$ of total MC moves~\cite{Frenkel2002}. The specific symmetry of these moves depends on the microphase morphology that is simulated: for lamellae, the directions perpendicular to the lamellae fluctuate together, independently of the third; for cylinders, the ratio between the directions parallel to the cylinder is kept fixed, in order to preserve the structure; for the other phases, the cubic symmetry of the box is maintained. For the FCC-cluster crystal phase, sampling is also accelerated by a cluster volume move algorithm specifically designed for this phase. The move consists of \zy{simultaneously} changing the volume of the equilibrium lattice sites and the \zy{distance between the particles in the} cluster, \zy{in order to minimize the generation of overlaps}. Details can be found in Ref.~\onlinecite{Zhuang2016}.

The second TI step, which brings finite-density hard spheres in a field to fully-interacting SWL particles, follows a linear alchemical transformation of the total system energy with $\alpha\in[0,1]$
\begin{widetext}
\begin{equation*}
U_\alpha(\textbf{r}^N;\mathcal{F})=-(1-\alpha)\sum_{i=1}^{N}\mathcal{F}(\textbf{r}_{i})+\sum_{i=j+1}^N\sum_{j=1}^N [u_{\mathrm{HS}}(r_{ij})+\alpha \;u_{\mathrm{SALR}}(r_{ij})],
\end{equation*}
\end{widetext}
which gives\zy{~\cite{Frenkel2002}}
\begin{equation}
  f_\mathrm{c}(\beta,\rho)=f_\mathcal{F}(\beta,\rho) + \frac{1}{N}\int_{0}^{1}\ud\alpha
\left\langle\frac{\partial U_\alpha}{\partial \alpha} \right\rangle_\alpha.
\end{equation}
In order to minimize the \zy{error due to} numerical integration, a Gauss--Lobatto quadrature with $20$ points is used.

Summing the various free energy contributions along the integration path \zy{finally} gives
\begin{widetext}
\begin{align}
  f_\mathrm{c}(\beta,\rho)=&f^{\mathrm{id}}(\beta,\rho)+\int_Ve^{-\beta \mathcal{F}(\br)}\ud\br +\int_0^{\rho}\frac{P(\rho')-\rho'/\beta}{\rho'^2}\ud\rho \nonumber+\frac{1}{N}\int_{0}^{1}\ud\delta
  \left<\frac{\partial U_\delta}{\partial \delta} \right>_\delta,
\end{align}
\end{widetext}
which is the free energy for a system constrained to a given microphase morphology and lattice occupancy. The equilibrium free energy for the system, $f(\beta,\rho)$, is then obtained by minimizing the results with respect to $n_\mathrm{c}$.

\subsection{Cluster liquid regime}
Upon increasing the fluid density from the ideal gas limit, the system forms a fluid of clusters  before periodic microphases become thermodynamically stable. Clusters emerge suddenly yet continuously, and therefore the process does not formally \zy{correspond to} a phase transition. In order to quantify this emergence, we adapt the tools developed \zy{for studying} the critical micelle concentration (cmc) to \zy{which this process is analogous}~\cite{Zhuang2016}.


In the spirit of the classical definition of the cmc, which situates the crossover as the point of most abrupt change in the physicochemical properties of the system~\cite{Gelbart2012}, we \zy{situate the onset of clustering at} the minimum -- when it exists -- of
\begin{equation}
h(\rho)\equiv\frac{\beta P-\rho}{\rho^2},
\end{equation}
which measures deviations of the \zy{liquid} equation of state from \zy{that of an} ideal gas. The pronounced peak it exhibits in cluster-forming systems indicates a rapid change in the system properties.
In practice, we determine the cmc by first fitting the numerical results for the equation of state to
\begin{equation}\label{eq:cmcpressure}
\beta p=\rho_\mathrm{s}+\rho_\mathrm{m}+B_{2,\mathrm{ss}}\rho_\mathrm{s}^2+B_{2,\mathrm{sm}}\rho_\mathrm{s}\rho_\mathrm{m}+B_{2,\mathrm{mm}}\rho_\mathrm{m}^2,
\end{equation}
where $\rho_{\mathrm{s}}$ and $\rho_{\mathrm{m}}$  are the \zy{single--particle} and
cluster~(micelle) densities, respectively. Clusters are defined as containing at least two particles $ij$, \zy{in contact within their attractive range, i.e., $r_{ij}<\lambda$. Every other particle is deemed a monomer}. The second virial
coefficients $B_{2,\mathrm{mm}}$ and $B_{2,\mathrm{sm}}$ are fitted to the simulation results, while we calculate\zy{~\cite{Hansen1990}}
\begin{widetext}
\begin{equation}
B_{2,\mathrm{ss}} = \frac{2\pi}{3}\left[1-(\lambda^3-1)(e^{\beta}-1) - 6\int_{\lambda}^{\kappa}r^2\left(\mathrm{e}^{-\beta\xi(\kappa-r)}-1\right)\ud r\right].
\end{equation}
\end{widetext}

In the cluster fluid regime, a fraction $1/N$ of the MC moves are virtual moves~\cite{Whitelam2007}, which enable efficient cluster displacements. As these moves have a significant computational overhead, however, they are only used in the nonpercolated cluster fluid regime, where the increased sampling efficiency warrants the cost.

\subsection{Cluster percolation transition}
\zy{Over a reasonably large range of temperatures,} further increasing the fluid density, results in the roughly spherical clusters first turning into wormlike clusters and
then into a disordered percolating network. Here again, although this change does not correspond to a thermodynamic transition, it nonetheless marks a significant change in the MC sampling efficiency. Cluster fluid configurations are \zy{relatively straightforward to sample}~(\zy{see} above), but the percolated fluid requires the use of parallel tempering (see below). It is thus algorithmically useful to locate the onset of fluid percolation.

\begin{figure}
	\includegraphics[width=0.5\textwidth]{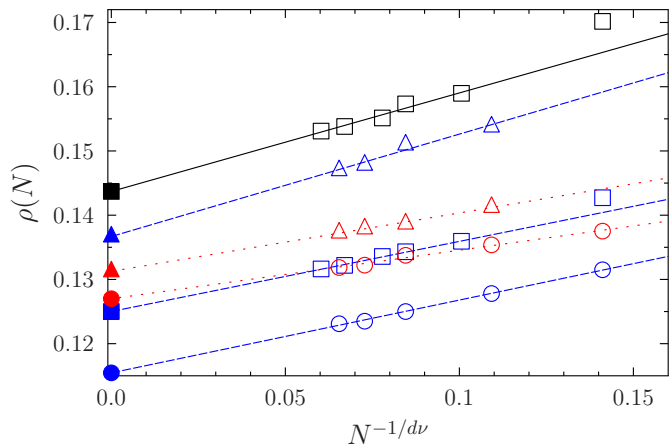}
	\caption{Extraction of the thermodynamic percolation transition, $\rho_{\mathrm{p}}(T)$, (solid \zy{markers}) by finite-size scaling for $T=0.60$ (black, \zy{solid}), $0.90$ (blue, \zy{dashed}) and $1.2$ (red, \zy{dotted}). Lines are fits of the simulation results to Eq. ~\ref{eq:perc} \zy{with SALR parameters (Eq. \ref{eq:swl})} $\kappa=2$ and $\xi=6.0$~(empty circles),  $\kappa=4$ and $\xi=0.05$~(empty squares), and $\kappa=6$ and $\xi=0.0318$ (empty triangles). \zy{Note that different colors represent different temperatures}.}\label{fig:percolation}
\end{figure}

\zy{Here again,} we define a pair $ij$ of particles to be in contact if they are within their attraction range, i.e., if $r_{ij}<\lambda$. The percolation transition, $\rho_{\mathrm{p}}(T)$, of the fluid is then determined by finite-size scaling of the midpoint of the percolation probability for different system sizes
\begin{equation}
\label{eq:perc}
\Delta \rho(N)\equiv\left|\rho_{\frac12}(N;T)-\rho_{\mathrm{p}}(T)\right| = N^{-\frac{1}{d\nu}},
\end{equation}
where $d\nu=2.706$ is the universal critical scaling for
three-dimensional standard percolation~\cite{Stauffer1994}. The intercept gives the percolation onset in the thermodynamic limit, $N\rightarrow\infty$(Fig.~\ref{fig:percolation}).

Sampling \zy{in this regime} is achieved \zy{by} using $2\times10^5$ MC steps of elementary AVB moves and random local displacements for $N=200$ to 1600. The percolation probability is obtained over at least $20$ configurations, and over all box directions, in order to minimize finite-size effects~\cite{Stauffer1994,Newman2000}.

\subsection{Percolated fluid regime}
In order to efficiently sample \zy{a configuration} within the percolated fluid regime, we use a standard parallel-tempering scheme with attempts to exchange configurations taking place a fraction $1/N$ of the MC moves. The temperature intervals and the number of configurations are chosen to ensure regular replica exchange takes place. For $\kappa=4$, for instance, we find that $\Delta T=0.0125$ for $T<0.55$ and $\Delta T=0.0250$ for $T>0.55$ suffice for systems with $N=600$ to $2400$. In order to sample ergodically, for $\rho<0.25$ the temperature chain goes up to $T=0.70$, while for $\rho>0.45$ the chain goes up to $T=1.00$. In total, $20$ to $100$ system replicas are thus simulated.

\begin{figure}
	\centering
	\includegraphics[width=0.5\textwidth]{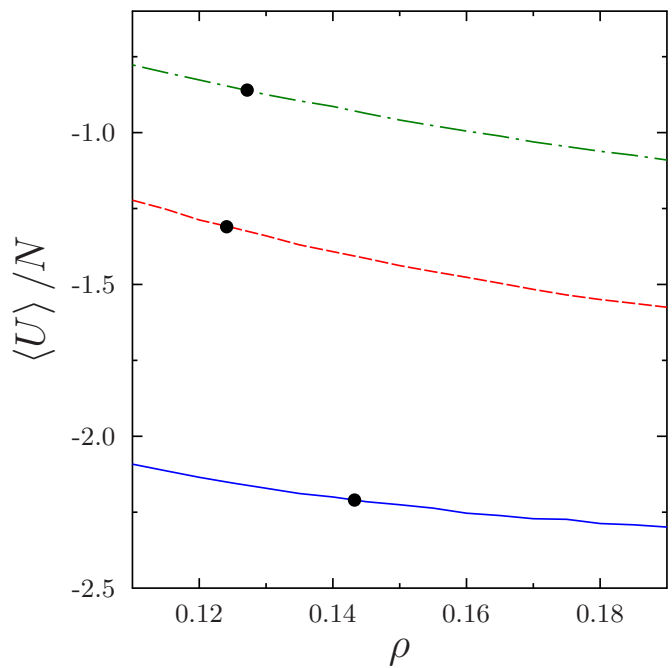}
	\caption{Average potential energy $\langle U\rangle$ of the disordered phase at $\kappa=4$, $\xi=0.05$ and $T=0.6$~(blue), $T=0.8$~(red) and $T=1.0$~(green) \zy{for $N=2000$}. Black markers denote the percolation transition. Note that the evolution of the energy is continuous and smooth over this entire regime.}
	\label{fig:percolation_rho_f}
\end{figure}

The resulting equilibrium configurations enable the free energy to be obtained by standard TI over $T$ and $\rho$~\cite{Frenkel2002}. Note that no discontinuity is observed in either the average potential energy $\langle U\rangle$ (Fig. \ref{fig:percolation_rho_f}) or the free energy (not shown), which suggests that a good sampling is obtained. This is also consistent with the absence of a phase transition \zy{around the percolation threshold as expected}.

\subsection{Gas-liquid coexistence, Lifshitz point and order--disorder transition}
As mentioned in the Introduction, microphases form when the long-range repulsion is sufficiently strong, i.e. when $\xi$ is above the Lifshitz point, $\xi_{\mathrm{L}}(\kappa)$. For $\xi<\xi_{\mathrm{L}}(\kappa)$, a standard gas-liquid coexistence is observed instead.
In order to set a lower bound on $\xi_{\mathrm{L}}(\kappa)$, we determine the gas-liquid coexistence regime of small $\xi$ systems using a Gibbs Ensemble Monte Carlo (GEMC) scheme~\cite{Frenkel2002}. We simulate systems with $N=512$ for $\kappa=$2 to 4, and with $N=1024$ for $\kappa=5$ and 6, using standard GEMC moves~\cite{Frenkel2002}: local single-particle displacements, volume exchanges for a fraction $1/N$ of the moves, and particle exchanges for $10\%$ of the moves.

The critical point, $(T_{\mathrm{c}},\rho_{\mathrm{c}})$, is estimated from fitting the coexistence results at various temperatures
\begin{equation}
\label{eq:crit}
\rho_{\pm} = \rho_{\mathrm{c}}+2C_2\left|1-\frac{T}{T_\mathrm{c}}\right|\pm \frac{1}{2}B_0\left|1-\frac{T}{T_\mathrm{c}} \right|^{\beta_{\mathrm{c}}},
\end{equation}
where $C_2$ and $B_0$ are fit parameters and $\beta_{\mathrm{c}}=0.3264$ is the three-dimensional Ising universality class critical exponent\zy{~\cite{Yeomans1992}}.

\begin{figure}
	\centering
	\includegraphics[width=0.5\textwidth]{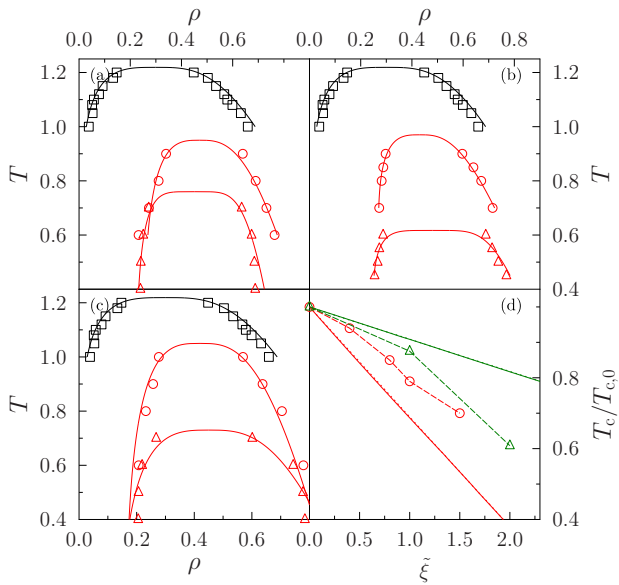}
	\caption{GEMC results for the gas-liquid coexistence density for the pure square--well model (squares) and for (a) $\kappa=2$ with $\xi=1.0$~(circles) and $3.0$~(triangles), (b) $\kappa=4$ with $\xi=0.01$~(circles) and $0.02$~(triangles), and (c) $\kappa=6$ with $\xi=0.001$~(circles) and $0.002$~(triangles). \zy{Increasing $T$ steadily depresses the critical point and narrows the coexistence binodal}. (d) Scaling of $T_c$ normalized by its $\xi=0$ value, $T_{\mathrm{c},0}$, for $\kappa=2$ (circles) with $\tilde{\xi}=\xi$, 
          and $\kappa=6$ (triangles) with $\tilde{\xi}=1000\xi$ from simulations (dashed lines, $T_{\mathrm{c},0}=1.20$) from RPA (solid lines, $T_{\mathrm{c},0}^{\mathrm{RPA}} =1.267$) and from DFT (dotted lines, $T_{\mathrm{c},0}^{\mathrm{DFT}}=1.395$). Note that on this scale the RPA and DFT predictions are nearly indistinguishable.
          \zy{Although both theories predict a depression of the critical point with increasing $\xi$, the sign of the discrepancy between theory and simulations changes with $\kappa$.}}
	\label{fig:coexistence}
\end{figure}

Figure \ref{fig:coexistence}a--c shows the simulation results for $\kappa=2$, $4$ and $6$. From the failure of the GEMC coexistence determination, we estimate $\xi_\mathrm{L}(2)=4.5(5)$, $\xi_\mathrm{L}(4)=0.025(5)$ and $\xi_\mathrm{L}(6)=0.0025(5)$. Note that the models we consider below have $\xi=6.0$ for $\kappa=2$, $\xi=0.05$ for $\kappa=4$, and $\xi=0.0318$ for $\kappa=6$, which are all well above $\xi_\mathrm{L}(\kappa)$. \zy{These models are thus expected to form microphases.}


\begin{figure}
	\centering
	\includegraphics[width=0.5\textwidth]{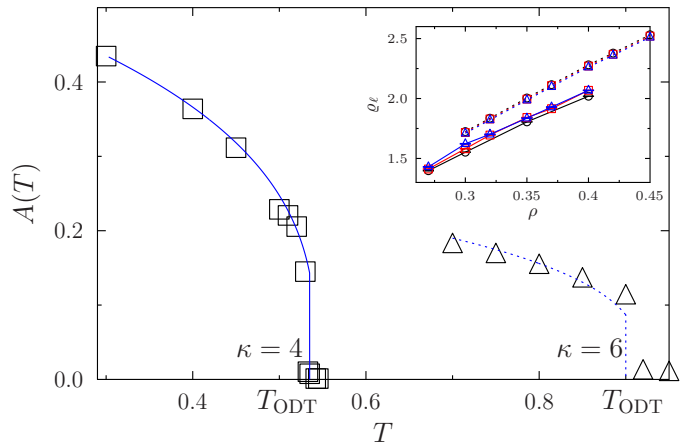}
	\caption{
		 Decay of $A(T)$ for $N=8000$ at $\rho=0.35$ for $\kappa=4$~(solid) and at $\rho=0.37$ for $\kappa=6$~(dashed) fitted to an Ising critical form $A(T)\sim\left|1-T/T_\mathrm{c}\right|^{\beta_{\mathrm{c}}}$ with $\beta_{\mathrm{c}}=0.3264$. \zy{The decay form and the discontinuity around $T_{\mathrm{ODT}}$ are consistent with the weakly first--order transition scenario.} We estimate $T_{\mathrm{ODT}}=0.535(5)$ for $\kappa=4$ and $T_{\mathrm{ODT}}=0.90(2)$ for $\kappa=6$. (Inset) Equilibrium occupancy of the lamellar phase at $\kappa=4$ and $6$ after minimizing the lattice occupancy for $T=0.3$ (dots),
		$T=0.35$ (squares) and $T=0.4$ (triangles) of $\kappa=4$~(solid line) and $\kappa=6$~(dashed line). Note that the equilibrium occupancy $\varrho_{\ell}$ is fairly independent of $T$ for both  $\kappa=4$ and $6$, \zy{which validates our protocol for measuring $A(k)$}.}
	\label{fig:ak}
\end{figure}

\zy{As described in the introduction,} for $\xi>\xi_{\mathrm{L}}$,  the second-order gas-liquid critical point is replaced by a weakly first-order, order-disorder transition (ODT)~\cite{Brazovskii1975,Bates1990,Seul1995}. The ODT is \zy{also} the highest temperature at
which periodic microphases melt. Because of the fairly symmetric form of the SWL interaction, the density where \zy{the ODT} takes place, $\rho_{\mathrm{ODT}}$, lies within the lamellar phase and maximizes their melting temperature~\cite{Ciach2013}. We can thus detect $T_{\mathrm{ODT}}$  by monitoring the decay of the lamellar order parameter
\begin{equation}
A(T)=\frac{1}{N}S(k_{\mathrm{c}};T).
\end{equation}
We denote $k_\mathrm{c}$ is the low-wavevector ($k_{\mathrm{c}}\in(0,2\pi)$) maximum of the structure factor
\begin{equation}
S(k;T)=\frac{1}{N}\left\langle\sum_{i\neq j} \mathrm{e}^{-i\mathbf{k}\cdot(\mathbf{r}_i-\mathbf{r}_j)}\right\rangle,
\end{equation}
at $\rho_{\mathrm{ODT}}$. Because $n_\mathrm{c}$ does not change much with $T$ within this regime (see inset of Fig.~\ref{fig:ak} \zy{for $n_\mathrm{c}$ at different $\rho$ and $T$}), the occupancy is kept constant over the course of these simulations. \zy{More specifically,} measurements are made with $N=8000$ and start from $\rho=0.37$ in the equilibrium lamellar phase with $\ell=5.23$ at $T=0.5$ for $\kappa=4$ and $\ell=5.72$ at $T=0.7$ for $\kappa=6$. The \zy{outcome of} steadily heating these configurations \zy{is reported in} in Figure~\ref{fig:ak}.

\section{Theoretical Descriptions}
\label{sec:theory}
In this section we consider two different theoretical estimates of the Lifshitz point for the SWL potential: (i) one based on the structure of the ground state, \zy{from a low--temperature approximation~(LTA) of the interaction energy}, and (ii) one based on the structure of the liquid, as approximated by the random phase approximation~(RPA). We also use RPA and a simple density-functional theory (DFT) to obtain the critical point, $T_\mathrm{c}(\kappa,\xi)$, for $\xi\leq\xi_{\mathrm{L}}(\kappa)$, and RPA alone for estimating the microphase envelope~($\lambda$--line) $\xi\leq\xi_{\mathrm{L}}(\kappa)$.

Note that a more elaborate DFT formulation could also be used to determine the $\lambda$--line and the microphase morphologies, as has been done in two dimensions.~\cite{Archer2008,Chacko2015} Extending this treatment to our three-dimensional system is, however, beyond the scope of the current work.

\subsection{Low-temperature approximation}
We first use LTA to estimate by considering the long length scale structure of the ground state. A compact ground state is assumed to result from the system undergoing a gas-liquid separation and hence to have a Ising-like critical point. By contrast, a ground state with mesoscale features is assumed to result from the system \zy{having undergone an ODT}. The transition from a compact to a mesoscale ground state should thus estimate $\xi_{\mathrm{L}}$.

In order to obtain a numerical estimate of the ground state structure, we follow the approach presented in Refs.~\onlinecite{Tarzia2006, Mossa2004}, and adapted to the case of particle microphase formers by Ciach~\cite{Ciach2013}.
In this scheme, mesoscale ordering is considered to emerge when a finite wavevector, i.e., $0<k<2\pi$ instead of $k=0$, dominates the Fourier transform of the average potential energy
\begin{equation}
\langle \tilde{U}\rangle(k;g(r))=\int\ud \br \exp\left(i \mathbf{k}\cdot\br\right){u_{\mathrm{SALR}}(r)g(r)},
\end{equation}
where $g(r)$ is the radial distribution function. 
A  minimum at $k>0$ can be found if $\langle \tilde{U}\rangle(0;g(r))$ is itself a maximum. The Lifshitz point that separates gas--liquid from microphase separation can thus be found by solving \zy{for $\xi$ in}
\begin{equation}
\left.\frac{\partial^2 \langle \tilde{U}\rangle(k;g(r))}{\partial k^2}\right|_{k=0}=0.
\end{equation}
Using a low-temperature and low-density approximation of the radial distribution function, i.e., $g(r)\approx\mathrm{e}^{-\beta u(r)}\approx\mathrm{e}^{-\beta u_{\mathrm{HS}}(r)}=\Theta(r-1)$, we obtain (for $\lambda=3/2$)
\begin{equation}\label{eq:LTA}
\xi_{\mathrm{L}}^{\mathrm{LTA}}(\kappa)= \frac{2532}{64\kappa^6-2916\kappa+3645}.
\end{equation}
The \zy{overall} result is depicted in Figure~\ref{fig:diff_xi} and, more specifically, $\xi_{\mathrm{L}}^{\mathrm{LTA}}(2)=1.32$, $\xi_{\mathrm{L}}^{\mathrm{LTA}}(4)=9.96\times10^{-3}$ and $\xi_{\mathrm{L}}^{\mathrm{LTA}}(6)=8.52\times10^{-4}$.

This approximation is clearly very crude, but improving the description of $g(r)$ is nontrivial. Although a better estimate of the HS structure could be considered, it would only slightly perturb the result because high-$k$ modifications do not carry over to the low-$k$ regime. \zy{The wavelength of interest is indeed much larger than the particle diameter}. Obtaining a better description of mesoscale ordering might \zy{more significantly} improve the estimate, but developing such a description is beyond the scope of our study. In the next subsection, we instead consider approaching microphase formation from the simple liquid regime, where structural approximations are more forthcoming.

\subsection{Random Phase Approximation}
\begin{figure}
	\centering
	\includegraphics[width=0.5\textwidth]{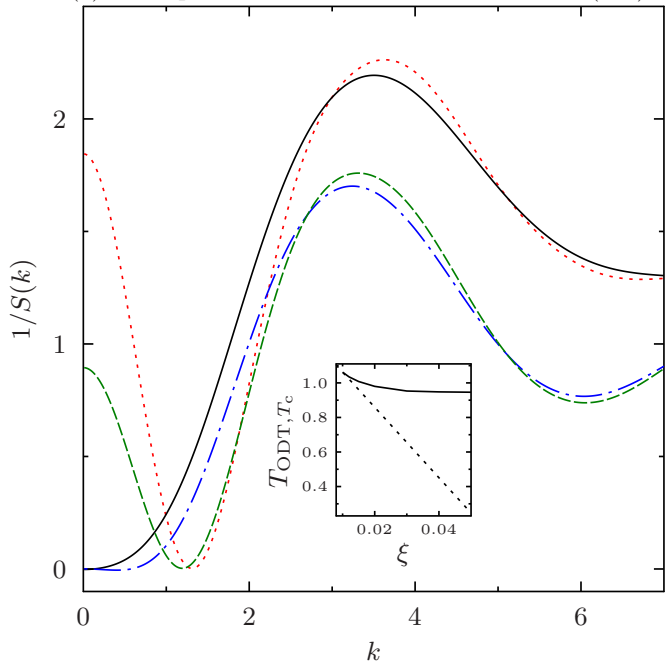}
	\caption{Highest $T$ divergence of the RPA structure factor for $\kappa=4$ at $\xi=0.008$ (solid, black) $\xi=0.01$~(blue, dash--dotted), $\xi=0.03$~(green, dashed), and $\xi=0.05$~(red, dotted). The structure factor diverges both at $T_\mathrm{c}$ and at $T_{\mathrm{ODT}}$, but the displacement of the divergence away from $k=0$ corresponds to the Lifshiz point, which for  $\kappa=4$ is $\xi_{\mathrm{L}}^{\mathrm{RPA}}=0.010(1)$. (Inset) Evolution of $T_{\mathrm{ODT}}$ (solid) and $T_{\mathrm{c}}$ (dashed) with $\xi$ for $\kappa=4$. For $\xi\approx\xi_{\mathrm{L}}^{\mathrm{RPA}}$, the highest temperature divergence occurs at $k>0$.}
	\label{fig:sk}
\end{figure}
\zy{A simillar line of argument gives that} a divergence of the structure factor, $S(k)$, \zy{from the liquid phase corresponds to} the homogeneous fluid phase becoming unstable to long length scale (small $k$) fluctuations. For a simple liquid, the gas-liquid spinodal can thus be extracted from the divergence of $S(0)$,\zy{ which is also the system compressibility $\rho k_{\mathrm{B}}T\chi=S(0)$~\cite{Hansen1986}}. A \zy{simillar} divergence \zy{of $S(k)$} at a small but non-zero $k$ indicates instead the emergence of clustering, which suggests at the formation of microphases.

The random phase approximation~(RPA) relates the structure factor to the Fourier transform of the Ornstein-Zernike direct correlation function $c(r;\rho)$,
\begin{equation}\label{eq:oz_relation}
	S(k)=\frac{1}{1-\rho \hat{c}(k;\rho)}.
\end{equation}
Following Ref.~\onlinecite{Archer2007}, here we treat the SALR contribution, $u_{\mathrm{SALR}}(r)$, as a perturbation of the Percus--Yevick~(PY) approximation for the structure of hard spheres. We thus obtain
\begin{equation}
	\hat{c}(k;\rho)=\hat{c}_{\mathrm{PY}}(k;\rho)-\beta  \hat{u}_{\mathrm{SALR}}(k),
\end{equation}
where
\begin{widetext}
\begin{equation}
		\hat{u}_{\mathrm{SALR}}(k)=-\frac{4\pi \left[2\xi\cos k\kappa
			+\left(-2\xi+k^2\lambda(\epsilon-\kappa\xi+\lambda\xi)
			\right)\cos k\lambda +k \left(\kappa\xi\sin
			k\kappa-(\epsilon-\kappa\xi+2\lambda\xi)\sin k\lambda \right)\right]}{k^4}.
\end{equation}
\end{widetext}
Recalling the PY approximation for three-dimensional HS~\cite{Ashcroft1966}
\begin{widetext}
\begin{align*}
	 \hat{c}_{\mathrm{PY}}(k;\rho)&=-4\pi\left[\left(\frac{\underline{\alpha}+2\underline{\beta}+4\underline{\gamma}}{k^3}-\frac{24\underline{\gamma}}{k^5}
	\right)\sin (k)\right.+\left(-\frac{\underline{\alpha}+\underline{\beta}+\underline{\gamma}}{k^2}+\frac{2\underline{\beta}+12\underline{\gamma}}{k^4}-\frac{24\underline{\gamma}}{k^6}
	\right)\cos (k)\left. +\left(\frac{24\underline{\gamma}}{k^6}-\frac{2\underline{\beta}}{k^4} \right)  \right],
\end{align*}
\end{widetext}
with $\underline{\alpha}=\frac{(1+2\eta)^2}{(1-\eta)^4}$, $\underline{\beta} =\frac{-6\eta(1+\eta/2)^2}{(1-\eta)^4}$ and 	 $\underline{\gamma}= \frac{\eta(1+2\eta)^2}{2(1-\eta)^4}$,
for the hard sphere volume fraction $\eta=\frac{1}{6}\pi\rho$, we can calculate the RPA estimate of $S(k)$ for the SWL potential. We then obtain
\begin{widetext}
\begin{equation}
	\rho k_{\mathrm{B}}T\chi=S(0)=\lim_{k\to 0}S(k) = \lim_{k\to 0} \frac{1}{1-\rho\left[ \hat{c}_{\mathrm{PY}}(k;\rho) - \beta \hat{u}_{\mathrm{SALR}}(k)
		\right]}
\end{equation}
\end{widetext}
with
\begin{equation*}
\begin{aligned}
&\lim_{k\to 0}\hat{c}_{\mathrm{PY}}(k;\rho) = \frac{\pi  \rho  \left(-1728 \rho ^3+72 \pi  \rho^2 -24 \pi ^2
	\rho  +\pi ^3 \right)}{\left(\pi -6 \rho
	\right)^4}\\
\mbox{and }&\lim_{k\to 0}\hat{u}_{\mathrm{SALR}}(k) = \frac{1}{3} \pi  \left[\xi  \left(\kappa ^4-4 \kappa  \lambda ^3+3 \lambda
^4\right)+4 \lambda ^3 \right].
\end{aligned}
\end{equation*}
From the two densities at which $\chi$ diverges, we identify the spinodal regime; the temperature at which the two branches meet at a single point gives ($T_{\mathrm{c}},\rho_{\mathrm{c}}$).

\begin{figure}
	\centering
	\includegraphics[width=0.5\textwidth]{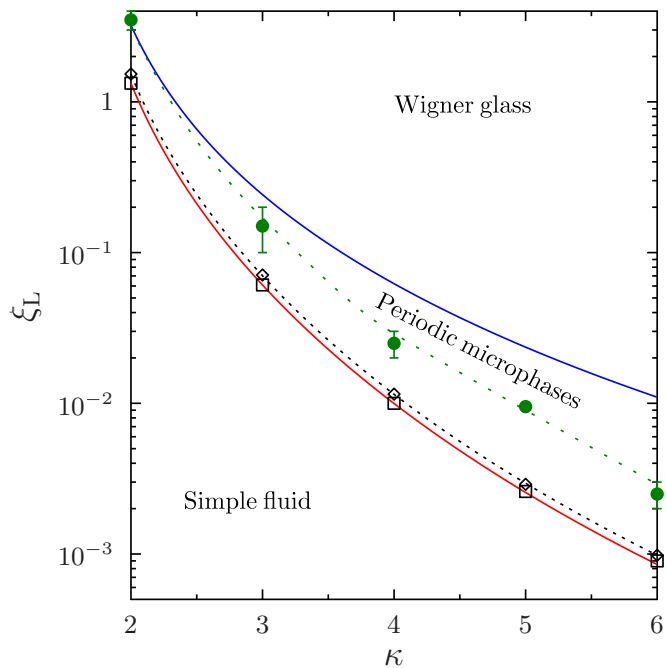}
	\caption{LTA (solid red line) and RPA \zyn{(full PY structure--empty squares; simple structure--empty diamonds; small $k$ expansion--dotted black line)} predictions for $\xi_{\mathrm{L}}$ along with GEMC simulation results~(green circles, dotted green line is a guide for the eye). RPA onset of the Wigner glass regime (solid blue line) obtained as described in the text. Note that the RPA and LTA results for $\xi_{\mathrm{L}}$ are nearly indistinguishable on this scale, but systematically underestimate the simulation results. \zy{Note also that the relative range of the periodic microphase regime shrinks with $\kappa$. Keeping $\xi$ away from the Wigner glass regime is thus expected to be easier at higher $\kappa$.}		}
	\label{fig:diff_xi}
\end{figure}

\zy{For $\xi>\xi_{\mathrm{L}}(\kappa)$, the gas-liquid spinodal and the critical point are replaced by the microphase regime and the order-disorder transition, respectively. The divergence of the structure factor then also takes place at $k>0$. The onset of the $k>0$ divergence thus determines the Lifshitz point, which can be extracted by expanding $S(k)$ at small $k$,
\begin{equation}
S(k)=S(0)-\gamma S(0)^2 k^2+\mathcal{O}(k^4),
\end{equation}
where
\begin{widetext}
\begin{equation}
  \gamma = \frac{1}{45} \pi  \rho  \left(-\beta  \left(\xi  \left(\kappa^6-6 \kappa  \lambda^5+5 \lambda^6\right)+6 \lambda^5 \varepsilon \right)-\frac{27 \left(2 \pi^2 \rho^2 -33 \pi  \rho +288\right)}{\left(\pi  \rho  -6\right)^4}\right).
\end{equation}
\end{widetext}
We obtain $\xi_{\mathrm{L}}=\frac{2916}{64\kappa^6-2916\kappa+3645}$ for $\lambda=3/2$, which is a form remarkably similar to the LTA result in Eq. \eqref{eq:LTA}, but with a numerator about $15\%$ larger. } We can also locate the boundary between $T_{\mathrm{ODT}}$ and $T_{\mathrm{c}}$ by examining the behavior of the full $S(k)$. The ODT occurs when $S(0)$ remains finite (and positive) as $S(k)$ diverges for $k=k_{\mathrm{c}}\in(0,2\pi)$. 
\zy{An example is provided in Figure~\ref{fig:sk}, and the results for $\xi_\mathrm{L}$ are given in Figure~\ref{fig:diff_xi}. Because the characteristic length of the first peak is much larger than the hard sphere diameter~\cite{Ciach2010}, it is not surprising to find that $\xi_{\mathrm{L}}(2)=1.33$, $\xi_{\mathrm{L}}(4)=0.010$, and $\xi_\mathrm{L}(6)=0.0009$ are fairly close to the low-$k$ expansion results $\xi_{\mathrm{L}}(2)=1.53$, $\xi_{\mathrm{L}}(4)=0.0115$, and $\xi_\mathrm{L}(6)=0.000981$}.

\zy{Generalizing this approach to different temperatures provides the pair of densities at which $S(0)$ remains finite and  positive while $S(k_\mathrm{c})$ for $k_\mathrm{c}\in(0,2\pi)$ diverges, and thus provides the $\lambda$--line. Here again, we can invoke the separation of length scales between microphase features and the particle diameter to simplify the hard sphere contribution to the structure as $\hat{c}_{\mathrm{PY}}(k;\rho)\approx \frac{(1+\pi\rho/3)^2}{\rho(1-\pi\rho/6)^4}$. 
The analysis of the divergence of $S(k)$ is then streamlined because the $k$ dependence comes from the SALR contribution alone. The wavevector that maximizes $S(k)$ thus solves $\left.\frac{\partial }{\partial k}\hat{u}_{\mathrm{SALR}}(k)\right|_{k=k_{\mathrm{c}}}=0$, and periodic microphases emerge when a solution occurs for $k_{\mathrm{c}}\in(0,2\pi)$. Note that this analysis also provides an estimate of $\xi_{\mathrm{L}}$, and, to order $k^2$ its value is identical to the low-$k$ expansion result.}

It has also been argued that for very strong repulsion, the formation of periodic microphases becomes impossible~\cite{Wu2004,Broccio2006,Sciortino2004}. For  $\hat{u}_{\mathrm{SALR}}(k)=0$, clusters are indeed hard-sphere-like, and for $\hat{u}_{\mathrm{SALR}}(k)>0$, cluster-cluster repulsion dominates. \zy{When repulsion dominates for all $k$,} the growth of larger mesoscale features is thus expected to be fully suppressed and the system to form a Wigner glass \zy{of clusters} at low temperatures. These conditions set an upper $\xi$ limit for the periodic microphase regime (Fig.~\ref{fig:diff_xi}).

\subsection{Simple density functional theory}
The isothermal compressibility can also be obtained from the liquid free energy, i.e., $\chi=\partial^2f/\partial v^2$. A simple DFT \zyn{(compared to the DFT of Ref.~\onlinecite{Archer2007})} of the SWL model can thus estimate the instability \zyn{of a uniform, disordered liquid}.
Assuming that the SWL liquid entropy is the same as that of a HS liquid, and that the SALR contribution is but a perturbation, the free energy \zyn{per particle of the constant-density disordered phase} becomes
\begin{widetext}
\begin{equation}
	f^{\mathrm{DFT}}(\beta,\rho)=f_{\mathrm{id}}(\rho)+f_{\mathrm{ex},\mathrm{HS}}(\rho)+\frac{N}{2V^2}\iint\ud
	\br\ud \br'v_{\mathrm{SALR}}(|\br-\br'|),
\end{equation}
\end{widetext}
where the PY approximation for HS gives
\begin{equation}
	\beta f_{\mathrm{ex},\mathrm{HS}}(\rho)=\left(\frac{3 \left(1-\frac{\eta }{2}\right)^2 \eta }{(1-\eta )^2}-\log (1-\eta )\right),
\end{equation}
and
\begin{equation*}
\iint\ud
	\br\ud \br'v_{\mathrm{SALR}}(|\br-\br'|)=\frac{\pi}{3} [\xi  (\kappa ^4-4 \kappa  \lambda ^3+3 \lambda ^4)+4 \lambda ^3].
\end{equation*}
The spinodal, however, only emerges for a specific range of repulsion strength, i.e., for small $\xi$, where normal gas--liquid phase separation takes place. We thus determine $T_{\mathrm{c}}$ by identifying where the two branches that delimit the onset of the unphysical regime, i.e., $\partial f/\partial v=0$, meet.

\section{Results and Discussion}
\label{sec:results}
\begin{figure}
	\centering
	\includegraphics[width=0.5\textwidth]{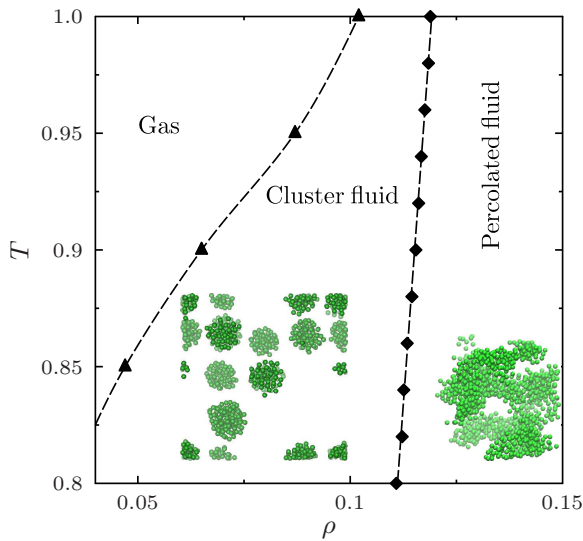}
	\caption{Summary $T-\rho$ phase diagram for $\kappa=2$ and $\xi=6$. The cmc (triangles) and percolation (diamonds) lines are depicted, but no gas-liquid or ordering phase transition could be detected. Lines are guides for the eye. The system is \zy{here} likely within the Wigner glass-forming regime of $\xi$.}
	\label{fig:pd_xi2}
\end{figure}

\begin{figure*}
	\centering
	\includegraphics[width=1\textwidth]{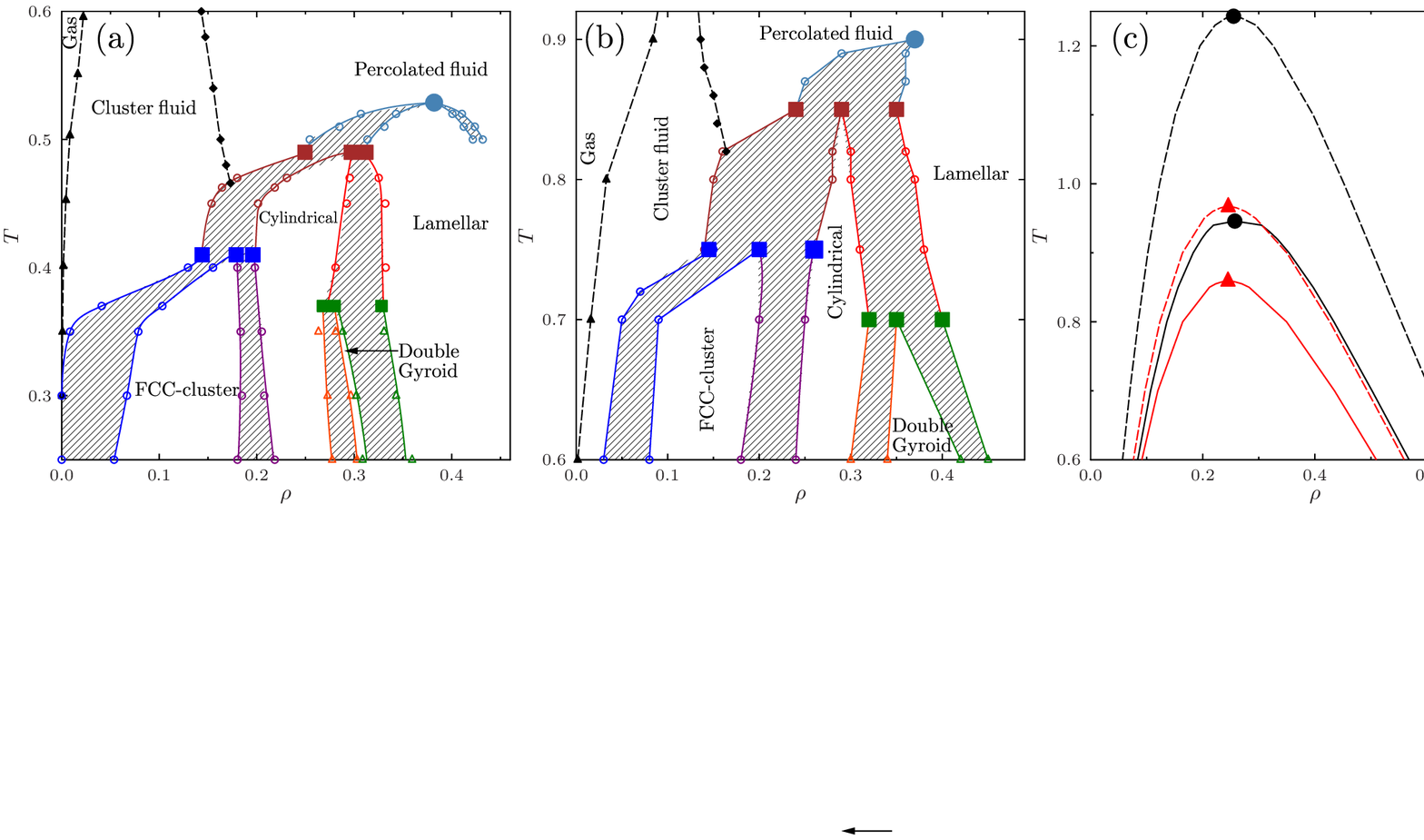}
	\caption{Summary $T-\rho$ phase diagram for (a) $\kappa=4$ and (b) $\kappa=6$. The cmc (triangles) and percolation (diamonds) lines modulate the disordered regime. The various ordered microphases found at intermediate densities delimit coexistence regimes (striped zones), triple points (squares) and the ODT (full circle). Lines are guides for the eye. (c) The RPA microphase envelope~(\zy{$\lambda$--line}) for $\kappa=4$ and $\xi=0.05$ (solid) and for $\kappa=6$ and $\xi=0.0318$~(dashed). Black circles denote $T_{\mathrm{ODT}}$ \zy{for RPA with the PY structure and red triangles the simpler HS structural.}}
	\label{fig:pd}
\end{figure*}
RPA and DFT predictions for $T_{\mathrm{c}}$ are obtained at weak $\xi$ (Fig.~\ref{fig:coexistence}d). As expected, $T_{\mathrm{c}}$ monotonically decreases with increasing $\xi$, and, after normalizing the results at $\xi=0$, the two approaches are essentially indistinguishable. This suggests a reasonably high degree of consistency between both mean-field-like description. The simulation results also follow qualitatively the same trend, but the \zy{actual numerical} estimates are relatively poor.

Both LTA and RPA yield predictions for $\xi_{\mathrm{L}}(\kappa)$~(Fig. \ref{fig:diff_xi}). This time, little quantitative difference is observed between the two approaches (only a few percent), despite them being based on markedly different physical principles. The two approaches are also systematically 2-3 times smaller than the numerical estimates of the Lifshitz point. This trend is consistent with the mean-field-like nature of the theories, \zy{and their neglect of fluctuations}. In a real three-dimensional system, repulsion must indeed be strengthened \zy{compared to the mean--field prediction} in order for periodic microphase order \zy{to spontaneously emerge}.

By contrast, the intermediate-temperature cluster fluid that is a precursor to the low-temperature Wigner glass is expected to be \zy{less} sensitive to fluctuations, \zy{because it} lacks long-range order. The periodic microphase regime being squeezed between the simple fluid and the Wigner glass regimes shrinks with $\kappa$ (Fig.~\ref{fig:diff_xi}). \zy{Fluctuations further} depress $\xi_{\mathrm{L}}(\kappa)$, \zy{and might} be sufficient for the ordered \zy{microphase} regime to disappear \zy{altogether} at small $\kappa$. For instance, \zy{for $\kappa = 2$}, $\xi_\mathrm{L}(2)$ is essentially superimposed with the RPA Wigner glass boundary, and despite our best \zy{numerical} efforts no stable periodic microphase regime could be located (Fig.~\ref{fig:pd_xi2}). The phase behavior of a system with  $\xi(2)=6>\xi_\mathrm{L}$ only displays a cmc-like clustering transition and the resulting clusters percolate upon further increasing density. The low-temperature regime remains disordered over all the $T$-$\rho$ range considered, the system gradually freezing \zy{in place as $T$ is lowered}. This phenomenon is akin to ``dynamic cluster glass'' formation~\cite{Wu2004,Broccio2006}. The quantitative prediction for the Wigner glass regime from RPA thus does seem physically reasonable for this model.

For $\kappa=4$ and $\kappa=6$, phase diagrams with periodic microphase regimes were obtained numerically (at $\xi=0.05$ and $\xi=0.0318$, respectively). Qualitatively, these two phase diagrams are fairly similar and  consistent with the periodic microphase sequence --
cluster crystal, cylindrical, double gyroid and lamellar phases --
of systems described by a comparable Landau free energy functional~\cite{Ciach2013}. (An exhaustive search for microphase morphologies was, however, not attempted~\cite{Zhuang2016}.) The main \zy{qualitative} differences are that the range of stability for the double gyroid \zy{is broader and the} microphase regime \zy{is more extended for $\kappa=6$}. The latter observation is consistent with the RPA predictions (Fig.~\ref{fig:pd}(c)). Indeed, although RPA overestimates the microphase envelope and $T_{\mathrm{ODT}}$, it does so consistently, predicting a broader and higher envelope for $\kappa=6$ than for $\kappa=4$. Remarkably, the discrepancy between RPA and the simulation results seems to shrink as the repulsion range increases ($T_{\mathrm{ODT}}^{\mathrm{RPA}}=0.946(1)$ v $T_{\mathrm{ODT}}=0.535(5)$ for $\kappa=4$ and $T_{\mathrm{ODT}}^{\mathrm{RPA}}=1.243(1)$ v $T_{\mathrm{ODT}}=0.90(2)$ for $\kappa=6$). A broader range of numerical results would, however, be needed to draw stronger conclusions from this observation. \zy{Surprisingly, the approximation that particles are uniformly distributed provides estimates closer to the numerical ODT than the PY approximation.}
	\zy{The improvement of the ODT estimate, however, does not mean that the neglect of the hard sphere structure provides a more physically-realistic estimate, but rather that the structural approximations used are somewhat crude and cannot be systematically improved. A final remark on RPA is that its predictions for $k_\mathrm{c}$ are fairly close to the numerical measurements, i.e., $k^{\mathrm{RPA}}_\mathrm{c}(4)=1.281$ and $k^{\mathrm{RPA}}_\mathrm{c}(6)=1.114$ vs $k_\mathrm{c}(4)=1.198$ and $k_\mathrm{c}(6)=1.112$. Interestingly, a similarly good agreement between mean-field treatments and numerical results was also observed in lattice systems~\cite{Zhang2011}.}

\zy{What may be most relevant for experiments is that} the interplay between the percolated and the cluster fluid regimes, on the one hand, and the ordered microphase regime, on the other, appears fairly independent of the model parameters for large enough $\kappa$. An equilibrium fluid of clusters and an equilibrium (gel-like) percolated fluid surround the ordered microphase regime, and should therefore \zy{generally} be taken into account when considering the assembly of \zy{periodic} microphases, \zy{both in simulation and in colloidal experiments}.

\section{Conclusion}
\label{sec:conclusion}
We have used a recently--developed TI-based simulation scheme for solving the phase diagram of a
continuous-space model with a microphase-forming regime.  Predictions from RPA and other theoretical approximations are shown to be in reasonably good qualitative agreement with the numerical results. Quantitative discrepancies can mostly be ascribed to the mean-field-like nature of the theoretical treatments. In particular, the methods predict a lower $\xi_{\mathrm{L}}$ than the simulation results, and increasing the repulsion range appears to improve the agreement.

The advance provided by the development of a numerical toolset for studying periodic microphases provides  complete coverage of the phenomenology exhibited by models with SALR. In particular, it reveals the rich thermodynamic interplay between order and disorder in these systems. We thus expect this approach to \zy{eventually} enable the complete elucidation of the dynamical behavior of \zy{colloidal microphase formers.~\cite{Seul1995, Campbell2005}}

\begin{acknowledgements}
We thank W.~Gelbart for his gracious support and encouragements over the years. We also thank K.~Zhang for help with early stages of this project. We acknowledge support from the National Science Foundation Grant No. NSF DMR-1055586. \zy{We also acknowledge support from the Materials Research Science and Engineering Centers~(DMR-1121107).}

\zy{Data relevant to this work have been archived and can be accessed at \url{http://dx.doi.org/10.7924/G8057CVF}.}
\end{acknowledgements}

\bibliography{./references}

\end{document}